\documentclass[ 12pt,onecolumn,notitle,reqno,fleqn,tbtags]{amsart}
\usepackage{amssymb,latexsym}
\begin{document}
\begin{center} Unified Analytic Electron Interaction Integrals Over Slater Orbitals for Diatomic Molecules \end{center}
 \begin{center} E.V.Rothstein \end{center}
\begin{center} evrothstein@gmail.com  \end{center}  

A unified formula for the analytical evaluation of two-center exchange, hybrid and coulombic type integrals is presented.
\begin{center} INTRODUCTION \end{center}
 A formula for the analytic evaluation of two-center electron interaction integrals is presented. 
 The terms in the sum  have been $ \text{checked}^{ 1 } $
\begin{center} FORMULA \end{center}

    \begin{multline*} \Phi_{ i c } ( j ) = ( - )^{ ( m_{ i } - | m_{ i } | ) / 2 }  ( 2 \delta_{ i } )^{ n_{ i } + \frac{ 1 }{ 2 } }   \left[ \frac{ ( 2 l_{ i } + 1 ) ( l_{ i } - | m_{ i } | ) ! }{ 4 \pi ( 2 n_{ i } ) ! ( l_{ i } + | m_{ i } | ) ! } \right]^{ \frac{ 1 }{ 2 } }    \\    \times    e^{ i m_{ i } \phi_{ j } } r_{ j c }^{ n_{ i } - 1 }  e^{ - \delta_{ i } r_{ j c } }  P^{ | m_{ i } | }_{ l_{ i } } ( cos \theta_{ j c } )    
   \end{multline*}   
       i = orbital number with quantum numbers $n_{ i} , l_{ i } , m_{ i } $ and screening constant  $ \delta_{ i } $ .  c = center, centered on nucleus a or b .        $\Phi_{ i c } ( j) $ = Normalized Slater Orbital for electron j, in spherical coordinates centered on center c.

  I = integral to be evaluated known as the two-center exchange integral      \begin{equation*} I = \langle \Phi_{ 1 a } ( 1 ) \Phi_{ 3 b } ( 1 ) \frac{ 1 }{ r_{ 1 2 } } \Phi_{ 2 a } ( 2 ) \Phi_{ 4 b } ( 2 ) \rangle  \end{equation*} 
 
\begin{multline} I = \delta ( \sum m_{ i } ) \delta ( | \sigma | - | m_{ 2 } + m_{ 4 } | )  W \sum_{ k } C_{ k } ( g_{ 1 k }, g_{ 2 k }, r_{ 1 k }, r_{ 2 k } )  \\ \times     \sum_{ \mu = | \sigma | } ( 2 \mu + 1 ) B( \mu, g_{ 1 k }, \beta_{ 1 } , | \sigma |) B( \mu, g_{ 2 k } , \beta_{ 2 } , | \sigma |) A( \mu, r_{ 1 k }, r_{ 2 k }, \alpha_{ 1 } , \alpha_{ 2 } , | \sigma |) \end{multline}
The $ \delta$ in [Eq. ( 1 )] is the Kronecker delta. The first sum is finite    over k ( the set of  $g_{ 1 k }, g_{ 2 K } , r_{ 1 k } , r_{ 2 k } $ generated by the four orbitals) .  The second sum is finite if either or both the $ \beta $ 's = 0 .  If $ \beta_{ 1 } = 0 $ the sum over $ \mu $ is restricted so the $ \mu + | \sigma | + g_{ 1 k } $ is even and the upper limit of the summation is  $ g_{ 1 k } +  | \sigma | $ .  If neither of the  $ \beta $'s = 0 the upper limit is infinite.  Some notation:  $P_{ l }^{ | m | } ( cos \theta ) $ = Associated  Legendre function of the first kind. ( see Ref. 2 , 8.6.6 )  R = internuclear distance between a and b nuclei.
\begin{multline*}   \alpha_{ 1 } = \frac{ R }{ 2 } ( \delta_{ 1 } + \delta_{ 3 } )  ; \quad  \alpha_{ 2 } = \frac{ R }{ 2 } ( \delta_{ 2 } + \delta_{ 4 } ) ;   \quad \beta_{ 1 } = \frac{ R }{ 2 } ( \delta_{ 1 } - \delta_{ 3 } ) ; \\  \quad      \beta_{ 2 } = \frac{ R }{ 2 } ( \delta_{ 2 } - \delta_{ 4 } ) ; \quad   W = R^{ \sum n_{ i } + 1 } \Pi \delta_{ i }^{ n_{ i } + 1 /2 } \\ W = \delta_{ 1 } ( \alpha_{ 1 } + \beta_{ 1 } )^{ n_{ 1 } - 1 / 2 } ( \alpha_{ 1 } - \beta_{ 1 } )^{ n_{ 3 } + 1/2 } ( \alpha_{ 2 } + \beta_{ 2 } )^{ n_{ 2 } + 1 / 2 } (\alpha_{ 2 } - \beta_{ 2 } )^{ n_{ 4 } + 1 / 2 } \\   W = \frac{ 1 }{ R } ( \alpha_{ 1 } + \beta_{ 1 } )^{ n_{ 1 } + 1 / 2 } ( \alpha_{ 1 } - \beta_{ 1 } )^{ n_{ 3 } + 1/2 } ( \alpha_{ 2 } + \beta_{ 2 } )^{ n_{ 2 } + 1 / 2 } (\alpha_{ 2 } - \beta_{ 2 } )^{ n_{ 4 } + 1 / 2 }                                                           \end{multline*}
\,  \,  \,  \,  \,  \,  \,  \,  \,  \,  \,  \,  \,  \,  \,  \,  \,\,  \,  \,  \,  \,  \,  \,  \,  \,  \,  \,  \,  \,  \,  \,  \,  \,  \,  \,  \,  \,  \,  \,  \,  \,  \,  \,  \,  \,  \,  \,  \,  \,  \,  \,  \,  \,  \,  \,  \,  \,  \,  \,  \,  \,  \,  \,  \,  \,  \,  \,  \,  \,  \,  \,  \,  \,  \,  \,  \,  \,  \,  \,  \,  \,  \,  \,  \,  \,  \,  \,  \,  \,  \,  \,  \,  \,  \,  \,  \,  \,  \,  \begin{multline}  \sum_{ k } C_{ k } ( g_{ 1 k }, g_{ 2 k }. r_{ 1 k } , r_{ 2 k } ) = ( - )^{ l_{ 3 } - | m_{ 3 } | + l_{ 4 } - | m_{ 4 } | }  \left( \frac{ 1 }{ 2 } \right)  \sum_{ a_{ 1 } = 0 }^{ n_{ 1 } - l_{ 1 } }  \sum_{ b_{ 1 } = 0 }^{ n_{ 3 } - l_{ 3 } } \\ \times   \sum_{ a_{ 2 } = 0 }^{ n_{ 2 } - l_{ 2 } } \sum_{ b_{ 2 } = 0 }^{ n_{ 4 } - l_{ 4 } }  ( -  )^{ b_{ 1 } + b_{ 2 } }  \binom{ n_{ 1 } - l_{ 1 } }{ a_{ 1 } }  \binom{ n_{ 3 } - l_{ 3 } }{ b_{ 1 } } \binom{ n_{ 2 } - l_{ 2 } }{ a_{ 2 } } \binom{ n_{ 4 } - l_{ 4 } }{ b_{ 2 } }  \\ \times \prod_{ i = 1 }^{ 4 } \left\{ \left[ \frac{ ( 2 l_{ i } + 1 ) ( l_{ i } - | m_{ i } | ) ! ( l_{ i } + | m_{ i } | ) ! }{ ( 2 n_{ i } ) ! } \right]^{ 1 / 2  } \,  \sum_{ s_{ i } = 0 }^{ [ ( l_{ i } - | m_{ i } | ) / 2 ] } \sum_{ p_{ i } = 0 }^{ 2 s_{ i } }    \right.     \\ \times       \sum_{ q_{ i } = 0 }^{ l_{ i } - | m_{ i } | - 2 s_{ i } }     \frac{ ( - )^{ s_{ i } + q_{ 3 } + q_{ 4 } + p_{ 3 } + p_{ 4 } }  }{ 2^{ l_{ i } } l_{ i } ! } \binom{2 l_{ i } - 2 s_{ i } }{ l_{ i } - | m_{ i } | - 2 s_{ i } } \binom{l_{ i } }{ s_{ i } } \\ \times   \binom{ l_{ i } - | m_{ i } | - 2 s_{ i } }{ q_{ i } }  \binom{ 2 s_{ i } }{ p_{ i } }          \}\  \sum_{ c_{ 1 } = 0 }^{ ( | m_{ 1 } | + | m_{ 3 } | - | \sigma | ) / 2 }  \sum_{ c_{ 2 } = 0 }^{ ( | m_{ 2 } | + | m_{ 4 } | - | \sigma | ) / 2 }        \\  \times       \sum_{ d_{ i } = 0 }^{ ( | m_{ 1 } | + | m_{ 3 } | - | \sigma | ) / 2 }         \sum_{ d_{ 2 } = 0 }^{ ( | m_{ 2 } | + | m_{ 4 } | - | \sigma | ) / 2 }   ( - )^{ c_{ 1 } + c_{ 2 } + d_{ 1 } + d_{ 2 } }   \\  \times             \binom{ ( | m_{ 1 } | + | m_{ 3 } | - | \sigma | ) / 2  }{ c_{ 1 } }  \binom{ ( | m_{ 1 } | + | m_{ 3 } | -| \sigma | ) / 2 }{ d_{ 1 } }              \binom{ ( | m_{ 2 } | + | m_{ 4 } | - | \sigma | ) / 2 }{ c_{ 2 } }  \\ \times         \binom{ ( | m_{ 2 } | + | m_{ 4 } | - | \sigma | ) / 2 }{ d_{ 2 } }    .                         \end{multline}                                                 The square brackets in the upper limit of summation over $s_{ i } $ is $ \frac{ 1 }{ 2 }  ( l_{ i } - | m_{ i } |  ) $  or   $ \frac{ 1 }{ 2 }  ( l_{ i } - | m_{ i }  |  - 1  ) $    whichever is an integral.                                 \begin{multline} g_{ 1 } = p_{ 1 } + p_{ 3 } + q_{ 1 } + q_{ 3 } + a_{ 1 } + b_{ 1 } + 2 c_{ 1 } , \\   g_{ 2 } = p_{ 2 } + p_{ 4 } + q_{ 2 } + q_{ 4 } + a_{ 2 } + b_{ 2 } + 2 c_{ 2 } , \\                                                     r_{ 1 } = 2 ( s_{ 1 } + s_{ 3 } + d_{ 1 } ) + q_{ 1 } + q_{ 3 } - p_{ 1 } - p_{ 3 } + n_{ 1 } + n_{ 3 } - l_{ 1 } - l_{ 3 } - a_{ 1 } - b_{ 1 }, \\             r_{ 2 } = 2 ( s_{ 2 } + s_{ 4 } + d_{ 2 } ) + q_{ 2 } + q_{ 4 } - p_{ 2 } - p_{ 4 } + n_{ 2 } + n_{ 4 } - l_{ 2 } - l_{ 4 } - a_{ 2 } - b_{ 2 }  , \text{ See footnote 3 } \\        B( \mu , g, \beta , | \sigma | ) = \left(  \frac { - \partial }{ \partial \beta } \right)^{ g } \sqrt{ 2 \pi } \,  \frac{ I_{ \mu + 1 / 2 } ( \beta ) }{ \beta^{ | \sigma | + 1 / 2 } }  , \\  I_{ \mu + \frac{ 1 }{ 2 }  } ( \beta ) = \text{modified spherical Bessel function of the first kind } \\ \text{   ( see 10.2.2 Ref. 2 ) } \\ \text{ If } \beta = 0 \text{  then,  }   B( \mu, g, \beta , | \sigma | ) =  \frac{ ( - )^{ \mu - | \sigma | } 2^{ \mu + 2 } \binom{ 1 + g + | \sigma | } {\frac{ g + | \sigma | - \mu }{ 2 } }  }{ ( 1 + | \sigma | ) ! \binom{ g + | \sigma | + 1 }{ g } \binom{ 2 + g + | \sigma | + \mu }{ 1 + \frac{ g + | \sigma | + \mu }{ 2 } }  }      \\                                \text{ If } \beta \neq  0 \text{  then,  }   B( \mu, g, \beta , | \sigma | ) =  \frac{ ( - )^{ \mu + 1 } }{ \beta^{ | \sigma | + 1 } }  \sum_{ k = 0 }^{ \mu } \frac{ \mu ! }{ ( \mu - k ) !}  \binom{ \mu + k }{ k }  \\ \times   \frac{ 1 }{ ( 2 \beta )^{ k } }   \sum_{ j = 0 }^{ g } \binom{ g }{ j } \binom{ k + | \sigma | + j }{ j } \frac{ j ! }{ \beta^{ j } }  \left[ ( - )^{ k + j + g + \mu + 1 } e^{ \beta }  + e^{ - \beta } \right]   \\  \text{ In practice this latter formula was not used.  The ascending } \\ \text{series formula proved to be more accurate ( less roundoff errors ) } \\  \text{ for } \beta \neq 0 \text{ and } 2 k + \mu - | \sigma | \geq g ,  \text{ then,   } \\   B( \mu , g, \beta , | \sigma | ) = ( - )^{ g } 2^{ \mu + 1 } \beta^{ \mu - | \sigma | - g }   \sum_{ k = 0 }^{ \infty } \frac{ \beta^{ 2 k } \binom{ \mu + 2 k - | \sigma | }{ g } \frac{ g ! }{ k ! \mu ! ( k + 1 ) ! } }{ \binom{ 2 \mu + 2 k + 1 }{ \mu + k } \binom{ \mu + k + 1 }{ \mu } }                                      \end{multline}              \begin{multline*}  A( \mu, r_{ 1  } ,  r_{ 2  } , \alpha_{ 1 } , \alpha_{ 2 } , | \sigma | ) =  \left( \frac{ - \partial}{ \partial \alpha_{ 1 } } \right)^{ r_{ 1 } }     \left( \frac{ - \partial}{ \partial \alpha_{ 2 } } \right)^{ r_{ 2 } }  \int_{ 1 }^{ \infty } \int_{ 1 }^{ \infty } d \xi_{ 1 } d \xi_{ 2 }   \\      \times  ( \xi_{ 1 }^{ 2 } - 1 )^{ | \sigma | / 2  }   ( \xi_{ 2 }^{ 2 } - 1 )^{ | \sigma | / 2  } P_{ \mu}^{ | \sigma | } ( \xi_{ 1 < 2 } ) Q_{ \mu }^{ | \sigma | } ( \xi_{ 2 > 1 } ) e^{ - \alpha_{ 1 } \xi_{ 1 } } e^{ - \alpha_{ 2 } \xi_{ 2 } }   \\   =  \frac{ e^{ - ( \alpha_{ 1 } + \alpha_{ 2 } ) } }{ 2^{ 2 \mu + 1 } }  \left[ \frac{ ( \mu + | \sigma | ) ! }{ \mu ! } \right]^{ 2 }  \sum_{ k = 0}^{ \left[ \frac{ \mu + | \sigma | }{ 2 } \right] }  ( - )^{ k } \binom{ 2 \mu - 2 k }{ \mu - | \sigma | } \binom{ \mu }{ k }  \\ \times    \sum_{ p = 0}^{ \left[ \frac{ \mu + | \sigma | }{ 2 } \right] }  ( - )^{ p } \binom{ 2 \mu - 2 p }{ \mu - | \sigma | } \binom{ \mu }{ p } \left\{   \right. \sum_{ n_{ 1 } = 0 }^{ k_{ 1 } } \frac{ k_{ 1 } ! }{ ( k_{ 1 } - n_{ 1 } ) !  \,  \alpha_{ 1 }^{ n_{ 1 } + 1 } }   \sum_{ n_{ 2 } = 0 }^{ k_{ 2 } } \frac{ k_{ 2 } ! }{ ( k_{ 2 } - n_{ 2 } ) !  \,  \alpha_{ 2 }^{ n_{ 2 } + 1 } } \\ \left[      \right.  \ln \left( \frac { 2 \alpha_{ 1 } \alpha_{ 2 } }{ \alpha_{ 1 } + \alpha_{ 2 } } \right)  + ( - )^{ n_{ 1 } + n_{ 2 } + k_{ 1 } + k_{ 2 } + 1 } e^{ 2 ( \alpha_{ 1 } + \alpha_{ 2 } ) }   E_{ 1 } ( 2 \alpha_{ 1 } + 2 \alpha_{ 2 } )  \\ + ( - )^{ k_{ 1 } + n_{ 1 } } e^{ 2 \alpha_{ 1 } }  E_{ 1 } ( 2 \alpha_{ 1 } )   +  ( - )^{ k_{ 2 } + n_{ 2 } } e^{ 2 \alpha_{ 2 } }  E_{ 1 } ( 2 \alpha_{ 2 } )  +  \gamma \left. \right] \\   + \sum_{ n_{ 2 } = 1 }^{ k_{ 2 } } \frac{ 1 }{ n_{ 2 } }  \sum_{ j_{ 2 } = 0 }^{ k_{ 2 } - n_{ 2 } } \frac{ k_{ 2 } !  }{ ( k_{ 2 } - n_{ 2 } - j_{ 2 } ) !  }   \sum_{ n_{ 1 } = 0 }^{ k_{ 1 } } \left[  \right.  \frac{ - k_{ 1 } ! }{ ( k_{ 1 } - n_{ 1 } ) ! }  \frac{   1  }{ \alpha_{1 }^{ n_{ 1 } + 1 }   \alpha_{ 2 }^{ n_{ 2 } + j_{ 2 } + 1 } }   \\ + \frac{  k_{ 1 } ! }{ ( k_{ 1 } - n_{ 1 } ) ! }   ( - )^{ k_{ 2 } + n_{ 2 } + j_{ 2 } } \sum_{ t = 0 }^{ n_{ 2 } - 1 } \frac{  2^{ n_{ 2 } - t - 1 } }{ ( n_{ 2 } - t - 1 ) ! \,  \alpha_{ 1 }^{ n_{ 1 } + 1 } \alpha_{ 2 }^{ j_{ 2 } + t + 2 } } \\ + \sum_{ j_{ 1 } = 0 }^{ n_{ 1 } }  \frac{ k_{ 1 } ! }{ ( n_{ 1 } - j_{ 1 } ) ! } \left(  \right.  \binom{ k_{ 1 } + n_{ 2 } - n_{ 1 } - 1 }{ k_{ 1 } - n_{ 1 } } \frac{  1 }{ \alpha_{ 2 }^{ j_{ 2 } + 1 } ( \alpha_{ 1 } + \alpha_{ 2 } )^{ k_{ 1 } + n_{ 2 } - n_{ 1 } } \alpha_{ 1 }^{ j_{ 1 } + 1 } }  \\ + \sum_{ t = 0 }^{ n_{ 2 } - 1 } \frac{ ( - )^{ k_{ 2 } + n_{ 2 } + j_{ 2 } + 1 } 2^{ n_{ 2 } - t - 1 } }{ ( n_{ 2 } - t - 1 ) ! }  \binom{ t + k_{ 1 } - n_{ 1 } }{ t }  \frac{  1 }{ \alpha_{ 2 }^{ j_{ 2 } + 1 } ( \alpha_{ 1 } + \alpha_{ 2 } )^{ k_{ 1 } + t + 1  - n_{ 1 } } \alpha_{ 1 }^{ j_{ 1 } + 1 } }   \left.  \right) \left.  \right]    \\   + \sum_{ n_{ 2 } = 0 }^{ k_{ 2 }  }  \frac{ k_{ 2 } ! }{ ( k_{ 2 } - n_{ 2 } ) ! } \sum_{ n_{ 1 } = 1 }^{ k_{ 1 } }  \frac{ 1 }{ n_{ 1 } } \sum_{ j_{ 1 } = 0 }^{  k_{ 1 } - n_{ 1 } } \frac{ k_{ 1 } ! }{ ( k_{ 1 } - n_{ 1 } - j_{ 1 } ) ! }  \left[   \right.   \frac{ 1 }{ \alpha_{ 1 }^{ j_{ 1 } + 1 } \alpha_{ 2 }^{ n_{ 2 } + 1 } ( \alpha_{ 1 } + \alpha_{ 2 } )^{ n_{ 1 } } }  \\ -  \frac{ 1 }{ \alpha_{ 1 }^{ n_{ 1 } + j_{ 1 } + 1 }  \alpha_{ 2 }^{ n_{ 2 } + 1 } }  + \sum_{ t = 0 }^{ n_{ 1 } - 1 } \frac{ ( - )^{ k_{ 1 } + j_{ 1 } + n_{ 1 } } 2^{ n_{ 1 } - 1 - t } }{ ( n_{ 1 } - t - 1 ) !  }   \left( \right.  \frac{ 1 }{ \alpha_{ 1 }^{ j_{ 1 } + t + 2 }  \alpha_{ 2 }^{ n_{ 2 } + 1 } }   \\ +  \frac{  ( - )^{ k_{ 2 } + n_{ 2 } + 1 } }{ ( \alpha_{ 1 } + \alpha_{ 2 } )^{ t + 1 } \alpha_{ 1 }^{ j_{ 1 } + 1 }  \alpha_{ 2 }^{ n_{ 2 } + 1 } }  \left. \right) \left.  \right] \left.  \right\}   +  \frac{ e^{ - ( \alpha_{ 1} + \alpha_{ 2 } ) }  ( \mu + | \sigma | ) !  }{ 2^{ 2  \mu }  \mu  !  }  \sum_{ k = 0 }^{ [ \frac{ \mu + | \sigma | }{ 2 } ] } ( - )^k \binom{ 2 \mu - 2 k }{ \mu - | \sigma | }   \\  \times   \binom{ \mu }{ k } \left[  \right.    \sum_{ \kappa  = 1 }^{ | \sigma | }  \frac{ ( - )^{ \kappa  }  | \sigma | !  }{ \kappa  } \binom{ \mu + | \sigma | - \kappa }{ \mu }  \sum_{ j = 0 }^{ [ \frac{ \kappa - 1 }{ 2 } ] } \binom{ \kappa }{ \kappa - 2 j - 1 }  \sum_{ n = 0 }^{ [ \frac{ \mu + | \sigma | - \kappa }{ 2 } ] } ( - )^{ n }    \\  \times    \binom{ 2 \mu - 2 n }{ \mu - | \sigma | + \kappa }  \binom{ \mu }{ n }        -  \sum_{ j = 0 }^{ [ \frac{ \mu - | \sigma | - 1 }{ 2 } ] }  \frac{ ( 2 \mu - 4 j - 1 ) ( \mu - 2 j - 1 + | \sigma | ) !  \, 2^{ 2 j + 1 } }{ ( 2 j + 1 ) ( \mu - j ) ( \mu - 2 j - 1 ) ! }   \\ \times \sum_{ n = 0 }^{ [ \frac{ \mu + | \sigma | - 2 j - 1 }{ 2 } ] }  ( - ) ^{ n } \binom{ 2 ( \mu - 2 j - 1 - n ) }{ \mu - 2 j - 1 - | \sigma | } \binom{ \mu - 2 j - 1 }{ n } \left.  \right]               \end{multline*}                                                                              \begin{multline}          \times \left[ \right.  \sum_{ n_{ 1 } = 0 }^{ k_{ 1 }^{ o } } \frac{ k_{ 1 }^{ o } ! }{ ( k_{ 1 }^{ o } - n_{ 1 } ) ! \, }  \sum_{ n_{ 2 } = 0 }^{ f_{ 2 } } \binom{ n_{ 1 } + f_{ 2 } - n_{ 2 } }{ f_{ 2 } - n_{ 2 } }   \\ \times       \sum_{ j_{ 2 } = 0 }^{ n_{ 2 } }  \frac{ f_{ 2 } !  }{ ( n_{ 2 } - j_{ 2 } ) ! \, \alpha_{ 2 }^{ j_{ 2 } + 1 } ( \alpha_{ 1 } + \alpha_{ 2 } )^{ f_{ 2 } + n_{ 1 } - n_{ 2 } + 1 } }   \\   + \sum_{ n_{ 2 } = 0 }^{ k_{ 2 } }  \frac{ k_{ 2 } ! }{ ( k_{ 2 } - n_{ 2 } ) ! \, }  \sum_{ n_{ 1 } = 0 }^{ f_{ 1 } } \binom{ n_{ 2 } + f_{ 1 } - n_{ 1 } }{ f_{ 1 } - n_{ 1 } }  \\ \times    \sum_{ j_{ 1 } = 0 }^{ n_{ 1 } }  \frac{ f_{ 1 } ! }{ ( n_{ 1 } - j_{ 1 } ) ! \, \alpha_{ 1 }^{ j_{ 1 } + 1 } ( \alpha_{ 1 } + \alpha_{ 2 } )^{ f_{ 1 } + n_{ 2 } - n_{ 1 } + 1 } }  \left.  \right]  .  \\  E_{ 1 } ( x )  =  \text{ Exponential integral  ( 5.1.1. Ref. 2 ) }  \\    \gamma = \text{ Euler's constant  ( table 1.1 Ref. 2 ) }   \\ k_{ 1 } = \mu + | \sigma | - 2 p + r_{ 1 }  , \quad  k_{ 2 } = \mu + | \sigma | - 2 k + r_{ 2 } , \\  k_{ 1 }^{ o } = \mu + | \sigma | - 2 k + r_{ 1 } , \quad  f_{ 1 } = \mu + | \sigma | - 2 ( n + j ) - 1 + r_{ 1 } , \\ f_{ 2 } = \mu + | \sigma | - 2 ( n + j ) - 1 + r_{ 2 }  ,                                            \end{multline}                                      1.  E.R.Chan, Chem. Phys. Lett., \underline{23} ,99 (1973) . \\                 2.  M. Abramowitz and I.A.Stegun, editors, \textit{Handbook of Mathematical Functions}, ( Dover, New York) \\                                                   3.   \begin{multline*}    Hybrid Integral = \langle \Phi_{ 1 a } ( 1 ) \Phi_{ 3 a } ( 1 ) \frac{ 1 }{ r_{ 1 2 } } \Phi_{ 2 a } ( 2 )  \Phi_{ 4 b } ( 2 ) \rangle  \\    \alpha_{ 1 } = \beta_{ 1 } = \frac{ R}{ 2 } ( \delta_{ 1 } + \delta_{ 3 } ) ;  \,  \alpha_{ 2 } = \frac{ R}{ 2 } ( \delta_{ 2 } + \delta_{ 4} ) ;  \,          \beta_{ 2 } =    \frac{ R}{ 2 } ( \delta_{ 2 } - \delta_{ 4 } ) ;  \\   Coulomb  Integral = \langle \Phi_{ 1 a } ( 1 ) \Phi_{ 3 a } ( 1 ) \frac{ 1 }{ r_{ 1 2 } } \Phi_{ 2 b } ( 2 )  \Phi_{ 4 b } ( 2 ) \rangle  \\    \alpha_{ 1 } = \beta_{ 1 } = \frac{ R}{ 2 } ( \delta_{ 1 } + \delta_{ 3 } ) ;  \,  \alpha_{ 2 } =  - \beta_{ 2 }  = \frac{ R}{ 2 } ( \delta_{ 2 } + \delta_{ 4} ) ;  \,                       \end{multline*}      These changes result in changes in the set of  $ g_{ 1 k } , g_{ 2 k }, r_{ 1 k }. r_{ 2 k }. C_{ k } $ .                                                                                 \end{document}